\documentclass[conference]{IEEEtran}
\IEEEoverridecommandlockouts
\usepackage{cite}
\usepackage{amsmath,amssymb,amsfonts}
\usepackage{algorithmic}
\usepackage{graphicx}
\usepackage{textcomp}
\usepackage{xcolor}
\usepackage{array}
\usepackage{cite}

\def\BibTeX{{\rm B\kern-.05em{\sc i\kern-.025em b}\kern-.08em
T\kern-.1667em\lower.7ex\hbox{E}\kern-.125emX}}

\begin{document}

\title{Numerical Investigations on Dilute Cold Plasma Potential and Electron Temperature}

\author{\IEEEauthorblockN{Shiying Cai, Chunpei Cai, Zhen Zhang}
\IEEEauthorblockA{\textit{Department of Mechanical Aerospace Engineering} \\
\textit{Michigan Technological University}\\
Houghton, MI 49931, USA \\
Email:ccai@mtu.edu}
}

\maketitle
\thispagestyle{plain}
\pagestyle{plain}

\begin{abstract}
Simulation results are presented to demonstrate electron temperature and electrical potential development in dilute and cold plasma development. The simulation method is a hybrid method which adopted fluid model for electrons due to their high mobility, while heavy ions and neutrals are modelled with the direct simulation Monte Carlo and Particle-In-Cell methods. The flows include steady, starting-up and shutting-down scenarios.  The goal is to illustrate the exponential behaviors which were predicted in several recently developed formulas. Those formulas include many coefficients related with local properties, and they are difficult to determine. Hence, those trends can only efficiently demonstrate by numerical simulations which are more convenient than experimental measurements. The results confirm several facts. For steady plasma flows, the temperature $T_e$ and potential profiles are smooth, very likely, they can be approximated with exponential functions. For unsteady flows, the property developing trends in the shutting down or starting-up processes change  monotonically. Further, at locations with large gradients, the property change trends are less ideal than those formulas. This is consistent with the assumptions with which those formulas were developed.  
\end{abstract}


\section{Introduction}
In dilute plasma flows, the continuum flow assumption breaks down, and there are multiple species, multiple time and space scales. For light electrons, the oscillation frequency and speeds are quite high, while heavy ions and neutrals have slow motions. For many situations, multiple physics must be considered, e.g., electric and magnetic fields, rarefication, and collisions.  Investigations on plasma flows are usually quite challenging, and many assumptions may be needed in order to perform theoretical, numerical and experimental studies.

Numerical simulations of dilute plasma are quite helpful to understand the related physics, and there are several simulation approaches. The first approach uses full scale Particle-In-Cell (PIC) \cite{PIC} simulations, where electrons, ions, and neutrals are treated as different individual particles. However, these full scale PIC simulations track each movement of each particle, with too many unnecessary details because for many situations, only the collective behaviors of different species are of special concerns. The simulation time steps in PIC simulations are limited by the periods of gyro-motion of electrons, ions, the mean collision time, and the plasma frequency. Due to these reasons, full-scale PIC simulations are quite expensive and not suitable to simulate many practical engineering plasma flow problems. The second approach solves the magneto-hydro-dynamic equations\cite{chen}, where ions and electrons are treated as a mixture. This method solves for the macroscopic properties, and it cannot offer separated information for electrons and ions. Another practical approach for plasma simulation is a hybrid method \cite{yim} which simulates heavy ions and neutrals with the direct simulation Monte Carlo (DSMC) \cite{DSMC} and the PIC methods; electrons are treated as fluids due to their higher mobility. For example, the fluid model is widely used to simulate plasma plume flows\cite{Review}. Compared with methods of solving ions and electron as  two fluids\cite{two}, this hybrid treatment provides detailed collective information for electrons and a certain degree of accuracy are maintained, while the simulation speed is much faster than full scale PIC simulations. In many plasma flow applications, the hybrid method is effective and sufficiently accurate. Other simulation approaches include the directly solving \cite{aristov} the Vlasov equation. However, numerical simulation results are usually case by case.

Experimental studies on plasma flows are important and sometimes un-replaceable because they can validate the theoretical and simulation results. However, related experiment costs can be high. By comparison, theoretical investigations and modelling can be helpful and can offer some deep insights.

For plasma flows, $T_e$ is one important property. Other than the electron number density and bulk velocity components, $T_e$ describes electron thermal motion behaviors and can reveal information for plasma flows. For example, dilute plasma flows from a Hall effect electric propulsion (EP) device can be used for thruster performance evaluations and contamination estimations, and $T_e$ can be measured conveniently. The electron thermal temperature and plasma potential increase  across the pre-sheath in front of a Tethered-Satellite-System (TSS) and it can result in significant current enhancement to the satellite\cite{Cooke}. It is also reported that new differential equations can be developed to describe the heat fluxes into the wall with two-temperature modelling of the thermal plasmas. Due to the significant temperature differences between the electrons and ions, usually electrons deposit significant energy into plasma \cite{pekker,weifz}.

Recent work illustrates modelling and analyzing $T_e$s and plasma potentials in both steady and unsteady dilute and cold plasma flows\cite{Cai0,Cai1,Cai2,unsteady}. The approach is based on energy equation for the fluid model for electrons\cite{Mitcher},
and the results are semi-analytical, while heavy ions and neutrals are not included. The number density gradients in the dilute plasma flows are assumed to be small, and the magnetic field effects are also assumed negligible.


Recently, the following relations on steady and unsteady plasma flows are obtain.  

\subsection{Steady state cor-relations}
\label{sec:unsteady1}
\begin{equation}
\begin{array}{rll}
    T_e(s)  &=& T_h - \frac{A_2}{A_1} +C_1  e^{\lambda_1 s} + C_2  e^{\lambda_2 s}    \\
            &\equiv& C_0 +C_1  e^{\lambda_1 s} + C_2 e^{\lambda_2 s};  \\
             \lambda_{1,2} &=& -\frac{A_{0s}}{2} \pm \frac{1}{2} \sqrt{A_{0s}^{2} - 4A_1}
\end{array}
\label{eqn:Tesoultion}
\end{equation}
where, $A_{0s}$ is the projected component of $\overrightarrow{A_0}$ along the stream line direction. Obviously, $\lambda_1$ is positive and $\lambda_2$ is negative; $C_0 = T_h - A_2/A_1$,  $C_1$ and $C_2$ are two constants to be determined with proper boundary conditions. For certain, along a streamline, one exponential function decays and the other increases. Evidently $\lambda_{1,2}$ values, and those coefficients  $C_0$, $C_1$, and$_2$ must  adjust along the streamline, to maintain bounded $T_e$. 

Correspondingly,  the plasma potential along a streamline is:
\begin{equation}
    \phi(s) =  \frac{k}{e} ( C_1  e^{\lambda_1 s}  +  C_2 e^{\lambda_2 s} ) + a_s s +b
\label{Eqn:potential}
\end{equation}
where  $a_s$ and $b$ are two constants to be determined with boundary conditions. Similar to the $T_e$ function, one exponential function decays and the other increases. 

\subsection{Unsteady state cor-relations}
\label{sec:unsteady}
In the above, two semi-analytical correlations for steady, dilute, cold plasma flows have been  obtained. One is for the plasma potential, and the other is for $T_e$. These two relations illustrate physical factors and their contribution to the steady electron flow properties. The relations hold along electron streamlines.   Those correlations are obtained with bold assumptions, which is unavoidable for modelling work.

If the steady flow and small density gradient assumptions are further relaxed,  it can be demonstrated that there still exist semi-analytical correlations for the plasma potential and $T_e$.

A solution to the $T_e$ is:
\begin{equation}
  T_e(s,t) = B_1 e^{D_6t} \left[ B_2 e^{r_1s}+B_3 e^{r_2s} \right]  - D_4 
\label{eqn:unsteady_Tesoultion}
\end{equation}
where $D_4$, $D_6$, $B_1$, $B_2$, $B_3$, $r_1$ and $r_2$ are constants to be determined. As shown, there may be oscillating wave behaviors in this distribution because  $r_1, r_2$ and $D_6$ can have imaginary components.  Eqns. \ref{eqn:unsteady_Tesoultion} and \ref{eqn:Tesoultion} are completely compatible, for steady flows, the time term in Eqn.\ref{eqn:unsteady_Tesoultion} disappears, and the equation automatically degenerates to Eqn. \ref{eqn:Tesoultion}.

The solution for the potential is:
\begin{equation}
\begin{array}{rll}
&&\phi(s,t) = L + \left( B_3 e^{-\sigma t/\epsilon_0}  + B_4  e^{D_6 t} \right) \frac{Y(s)}{K}  = L  \\
&& +\frac{1}{K} \left( B_3 e^{-\sigma t/\epsilon_0}  + B_4  e^{ D_6 t} \right) (B_2e^{r_1 s} + B_3e^{r_2 s})
\end{array}
\label{eqn:unstady_pot}
\end{equation}
where  $K$, $B_4$ and $L$  are constants to be determined. 

Equation \ref{eqn:unstady_pot}  indicates the potential develop with oscillations in time and locations along the electron streamline.  Because the term $\sigma/\epsilon >0$, the first term in the first bracket  decays with time. However, the other coefficients in this expression can include imaginary parts, the plasma potential can grow, decay, or remain stable for both time and streamline, and the oscillations with sine or cosine profiles are highly possible for time and space. -As well known, oscillations are fundamental features in plasma flows. For this situation, the Joule heating term can include oscillations in space and time as well, and the amplitude can vary, it effect is not limited to dissipation.  

\section{Numerical Simulation Results and Discussions}
This section presents the discussions for the above correlations between the $T_e$ and plasma potentials, both steady and unsteady flow situations.  These semi-analytical expressions are based on the energy equation for elections with the detailed fluid model, and along an electron streamline. It is difficult to validate them accurately due to  the  complex physics and those unknown  coefficients in the correlations.  However, it is convenient to observe whether there are the same trends with exponential functions or wave patterns with oscillations.  Three-dimensional numerical simulations of plasma plume flows from BHT200 and H6 EP devices with the hybrid method are performed, i.e. DSMC/PIC for heavy ions and neutrals, while the electron properties are obtained by adopting the detailed fluid model for $T_e$. The numerical solutions and correlations are expected to have the same trends, especially at locations without strong gradients.  Experimental studies are feasible, but they need special setups, and it may be inconvenient to identify specific curved streamlines along which properties shall be measured.

\subsection{Steady Flow Situation}
\label{sec:steady_validation}
For the steady flow situation, the primary concern in the validations is whether Eqn.\ref{eqn:Tesoultion} holds. To construct the temperature profile can be done conveniently with two steps. The first step is to sample the hybrid numerical simulation results at one specific point, based on the information, the local coefficients $A_1$, $A_2$, $\lambda_1$ and $\lambda_2$ can be computed. The second step is to choose the $T_e$ at another point along the same streamline. By using the temperature values at two points, the coefficients  $C_1$ and $C_2$ can be determined.

\begin{figure}[ht]
    \includegraphics[width=3.6in,height=2.9in]{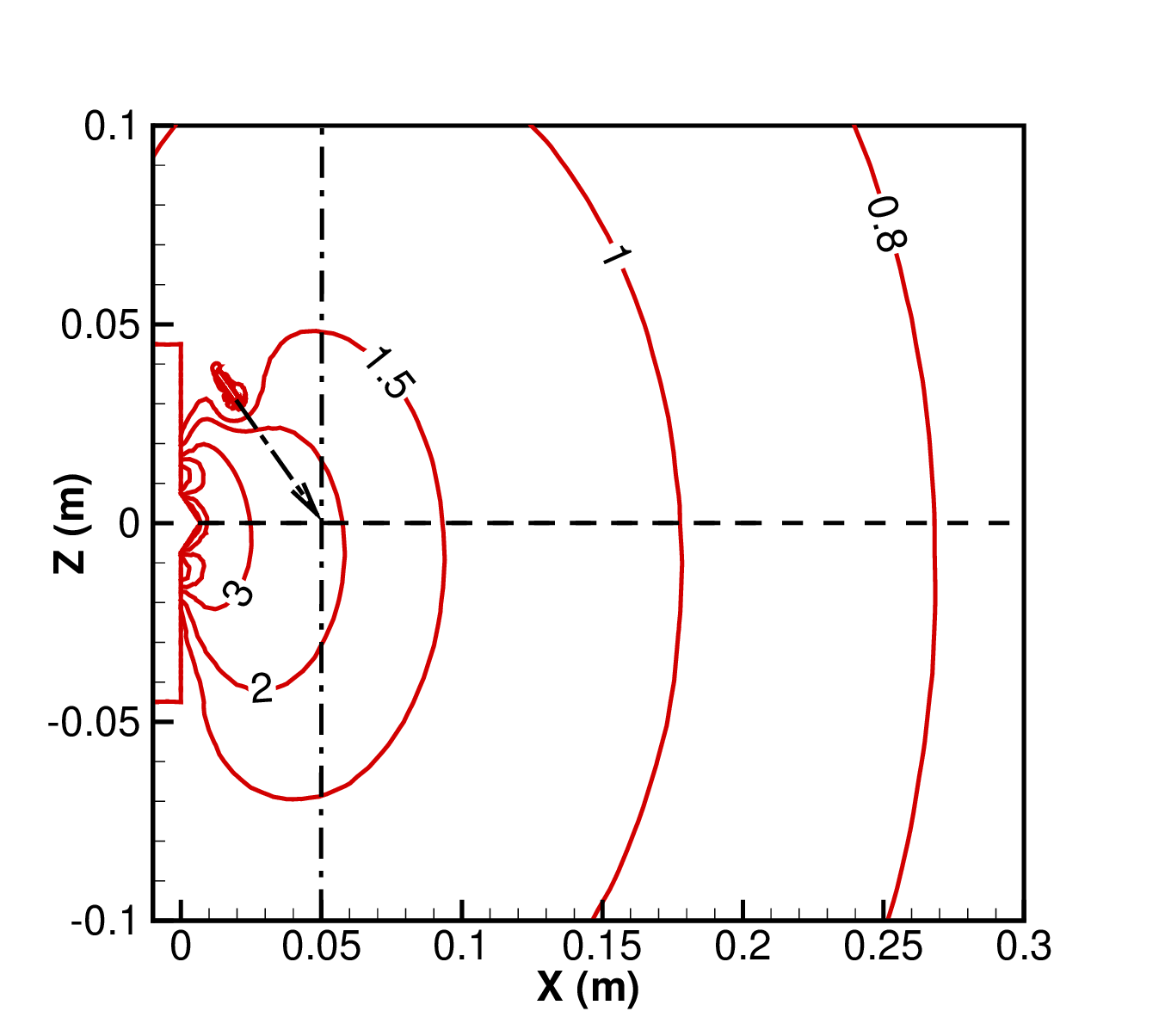}
      \caption{$T_e$ contours inside a symmetric plane from a cluster of two Hall thrusters, at a steady flow state. }
       \label{Fig:singleBHT_Te}
\end{figure}

In the first simulation, the magnetic leakage is assumed negligible outside the thruster acceleration channel. The detailed simulation parameters are available in the literature\cite{Cai0}. Figure \ref{Fig:singleBHT_Te} shows the $T_e$ contours in the XZ plane. The small triangular shape in front of the thruster center is a protection cap, and there is a cathode above the thruster. The cathode exit points to $X_P=0.05$ cm on the centerline. Electrons issued from the cathode impinge at the thruster centerline  at $X_p=0.05$ cm, which is the hottest spot in the figure. This $x_p$ divides the centerline into two regions.  The inner region  $0<x<x_p=0.05$ cm with large gradients, for example, the potential rises from 0.0 V at the thruster front face dramatically to the hottest spot. Obviously, in this region, some assumptions will not hold very well due to large gradients. In the outer region with $x>x_p=0.05$ cm. In the outer region with $x>x_p$, the gradients are mild and the derived formulas with embedded exponential functions may have better performance.  

\begin{figure}[ht]
    \includegraphics[width=3.6in,height=2.9in]{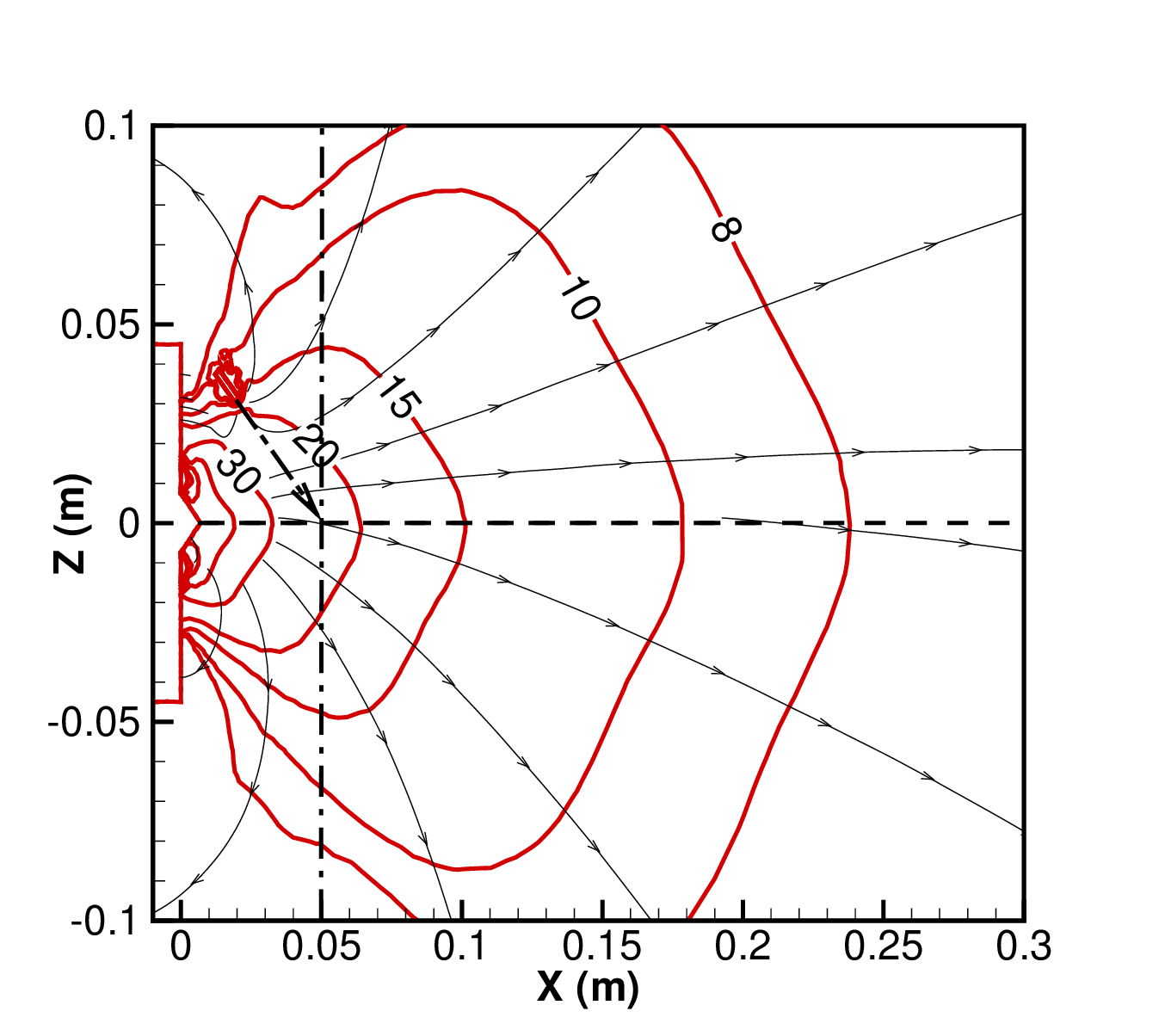}
        \caption{Potential contours inside a symmetric plane in a cluster of two Hall thrusters, at a steady flow state.}
        \label{Fig:singleBHT_phi}
\end{figure}

Figure \ref{Fig:singleBHT_phi} shows the corresponding potentials and streamlines inside the XZ plane. These two figures show similar patterns between the plasma potentials and $T_e$s inside the XZ plane. The properties along the thruster centerline are chosen for comparisons because there are some experimental measurements available at locations $X \le 0.2$ cm.  As illustrated by Fig.\ref{Fig:singleBHT_phi}, at locations with  $X \ge 0.2$ cm, the centerline can be approximated as a streamline. Another reason to adopt the centerline is because  $a_s$ is constant along a straight line;  however, along a curved streamline, $a_s$ is usually not constant. The streamlines in Fig. \ref{Fig:singleBHT_phi} illustrate that electrons flow out of the cathode and the thruster acceleration channel.

\begin{figure}[ht]
\includegraphics[width=3.6in,height=2.9in]{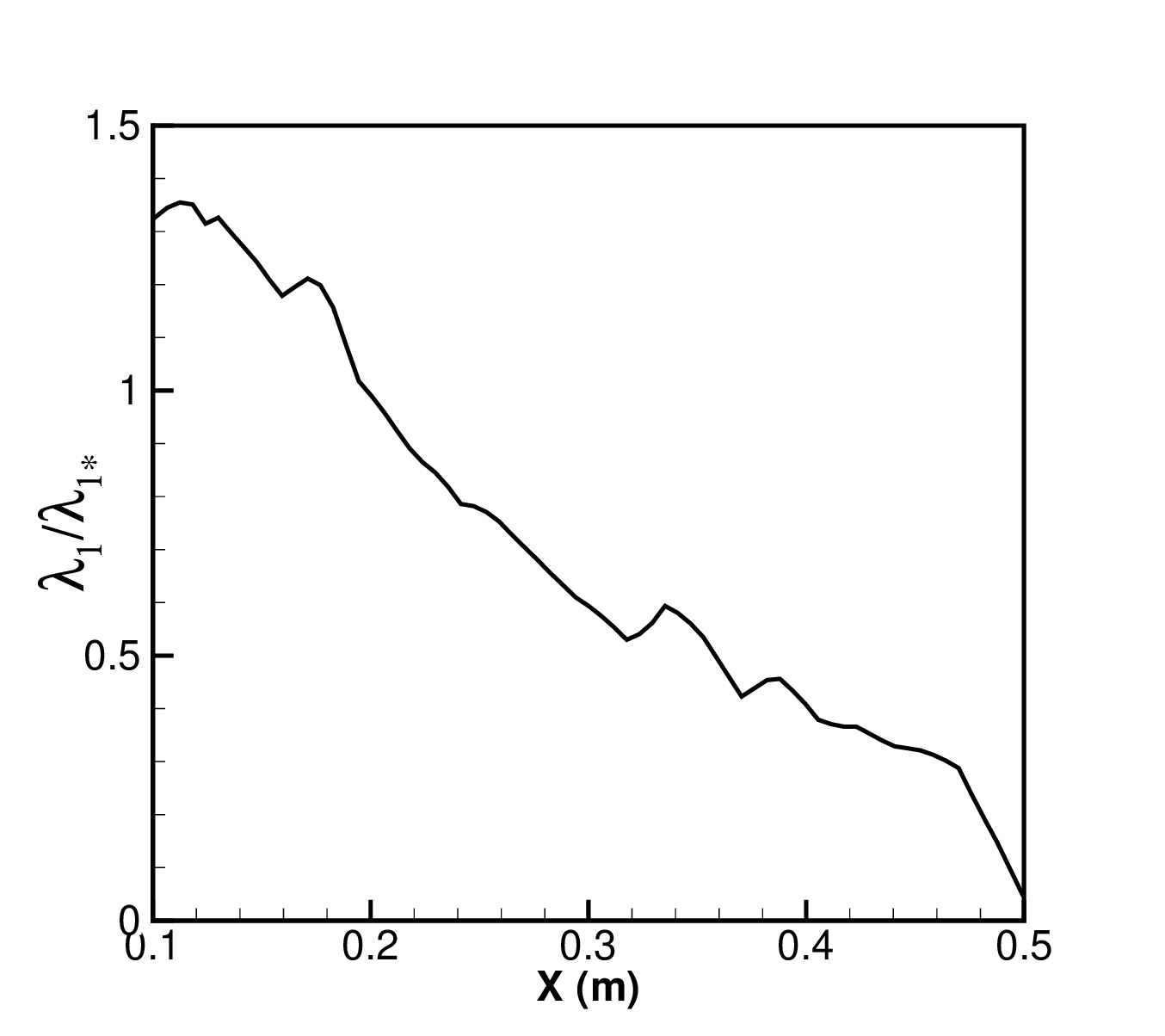}
      \caption{Normalized eigenvalue $\lambda_1$ distribution along a single BHT200 centerline.}
       \label{Fig:singleBHT_lambda1}
\end{figure}

\begin{figure}[ht]
    \includegraphics[width=3.6in,height=2.9in]{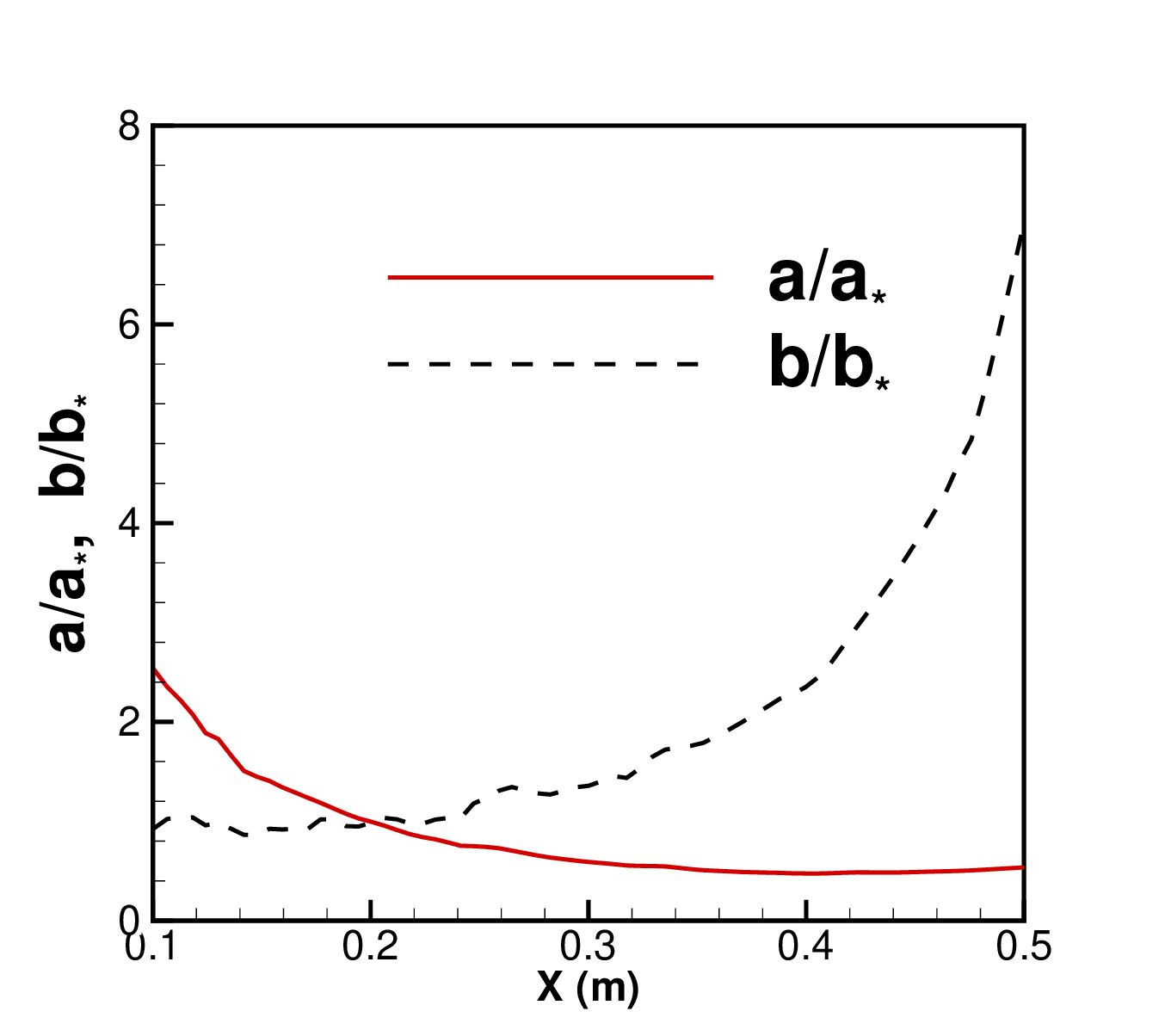}
      \caption{Normalized variables $a$ and $b$ distributions along a single BHT200 thruster centerline. }
      \label{Fig:singleBHT_ab}
\end{figure}

This new model for steady plasma flows assumes that local gradients are small and local properties are almost constant. This assumption can be tested by examining the hybrid simulation results. Figure  \ref{Fig:singleBHT_lambda1} shows the normalized centerline eigenvalue $\lambda_1$.  As illustrated by the last expression in Eqn.\ref{eqn:Tesoultion}, $\lambda_1$ is positive and $\lambda_2$ is negative.  The contribution from the term containing $\lambda_2$ to Eqn.\ref{eqn:Tesoultion} decreases quickly, and becomes negligible; hence, the corresponding figure is not included here.  Figure \ref{Fig:singleBHT_ab} shows the normalized $a_s$ and $b$ along the thruster centerline. The corresponding values at $X= 0.2$ cm are chosen as reference values for normalization. These two figures indicate within the distance range 0.2 cm $\le$ X $\le$ 0.4 cm, these three normalized parameters are in the order of unity; hence, it is acceptable to apply the new analytical results.

\begin{figure}[ht]
  \includegraphics[width=3.6in,height=2.9in]{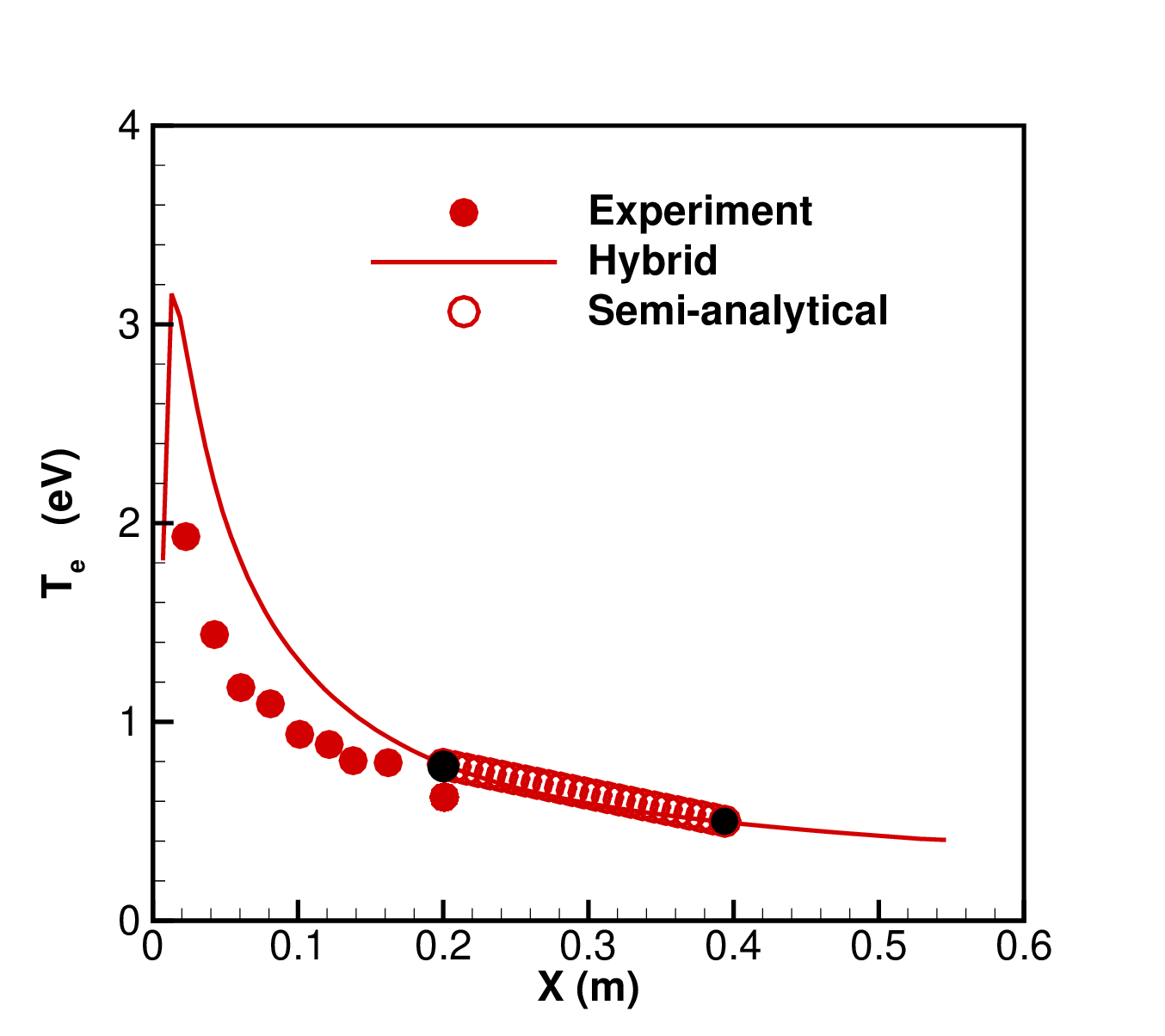}
      \caption{$T_e$ distribution along a BHT200 centerline. }
\label{Fig:singleBHT_cen_te_exp_detailed_fit}
\end{figure}

\begin{figure}[ht]
      \includegraphics[width=3.6in,height=2.9in]{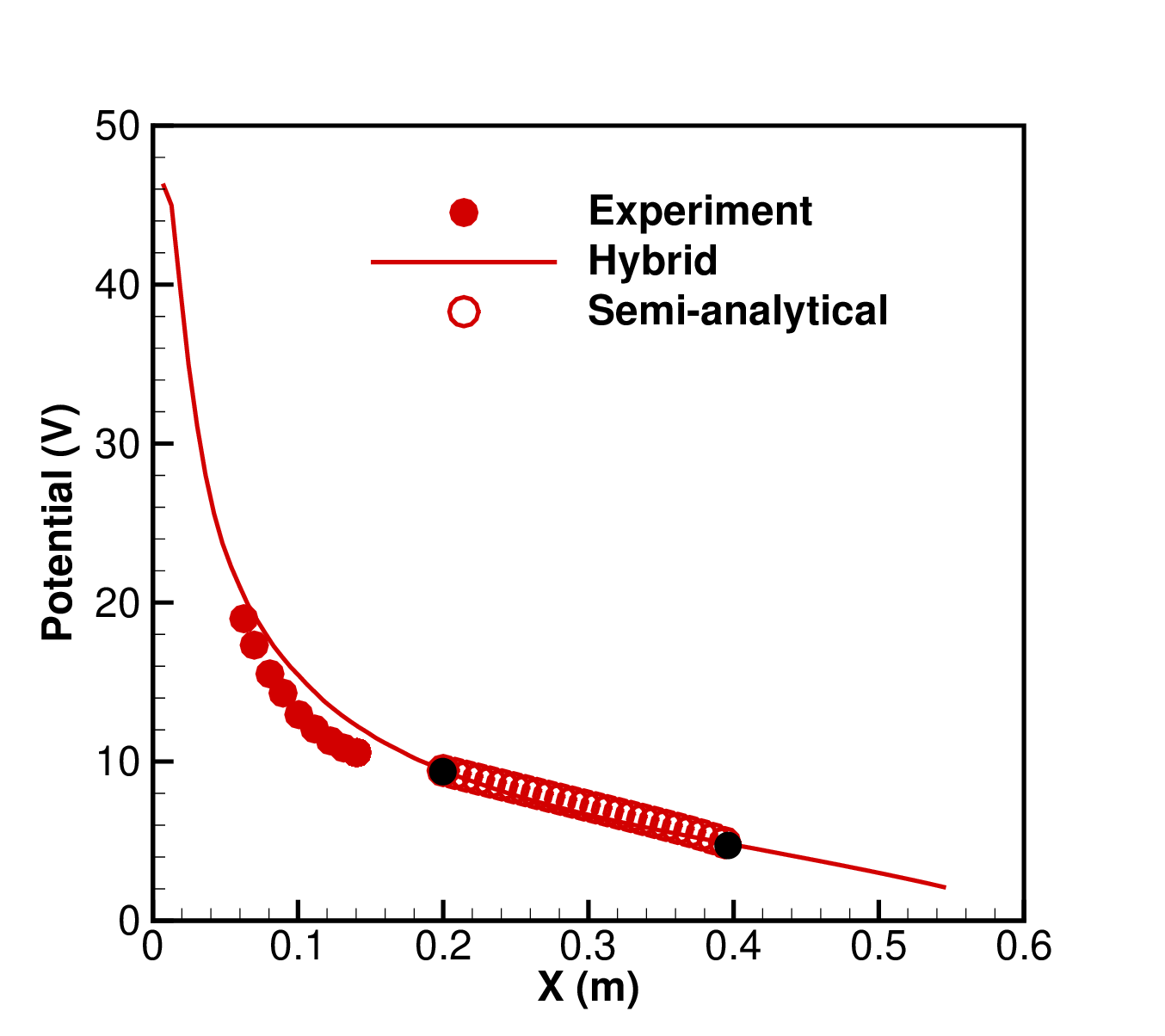}
      \caption{Plasma potential profile along a BHT200 thruster centerline.}
\label{Fig:singleBHT_cen_pot_exp_detailed_fit}
\end{figure}

Figure \ref{Fig:singleBHT_cen_te_exp_detailed_fit} illustrates the $T_e$ distributions along the BHT200 centerline, and Fig. \ref{Fig:singleBHT_cen_pot_exp_detailed_fit} illustrates the corresponding potential distributions. Both figures include experimental measurements at very near fields; the hybrid numerical simulation results; and the semi-analytical expressions by using the information at two points (marked with two black dots) from the numerical simulation results. The selections of these two points are relatively arbitrary, the main criteria is that they must be at locations with mild variations, which is one critical assumption to obtain Eqns.\ref{eqn:Tesoultion} and \ref{Eqn:potential}.  The semi-analytical results involve two exponential functions and one linear function, even though the profiles seem linear. The simulation results are reasonable because they are close to experimental measurements, and the constructed curves follow the same trends of the numerical simulation results.

\begin{figure}[ht]
      \includegraphics[width=3.6in,height=2.9in]{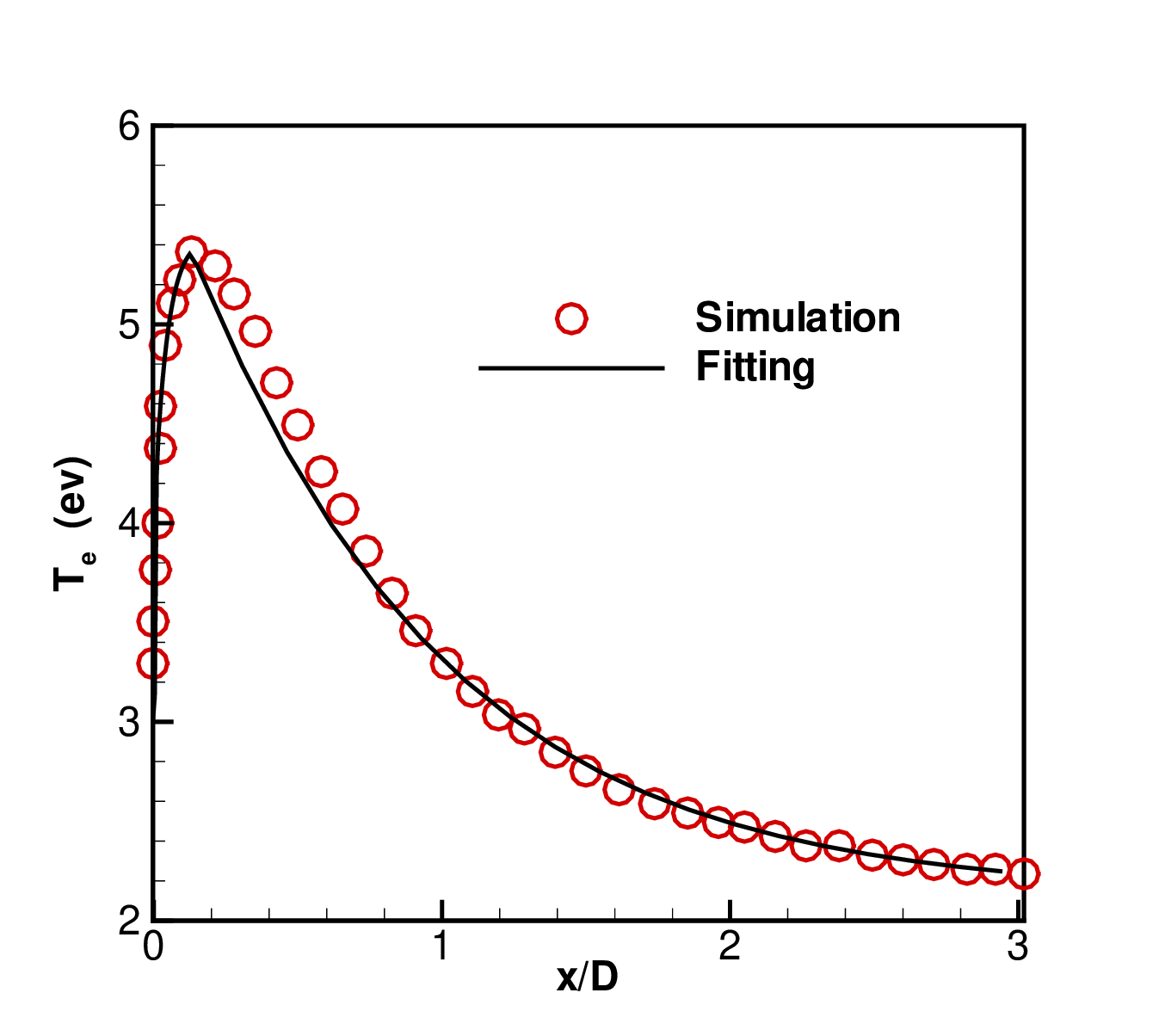}
      \caption{Curve-fitting results for $T_e$ profiles along an H6 centerline. }
       \label{Fig:H6_Te} 
\end{figure}

\begin{figure}[ht]
\includegraphics[width=3.6in,height=2.9in]{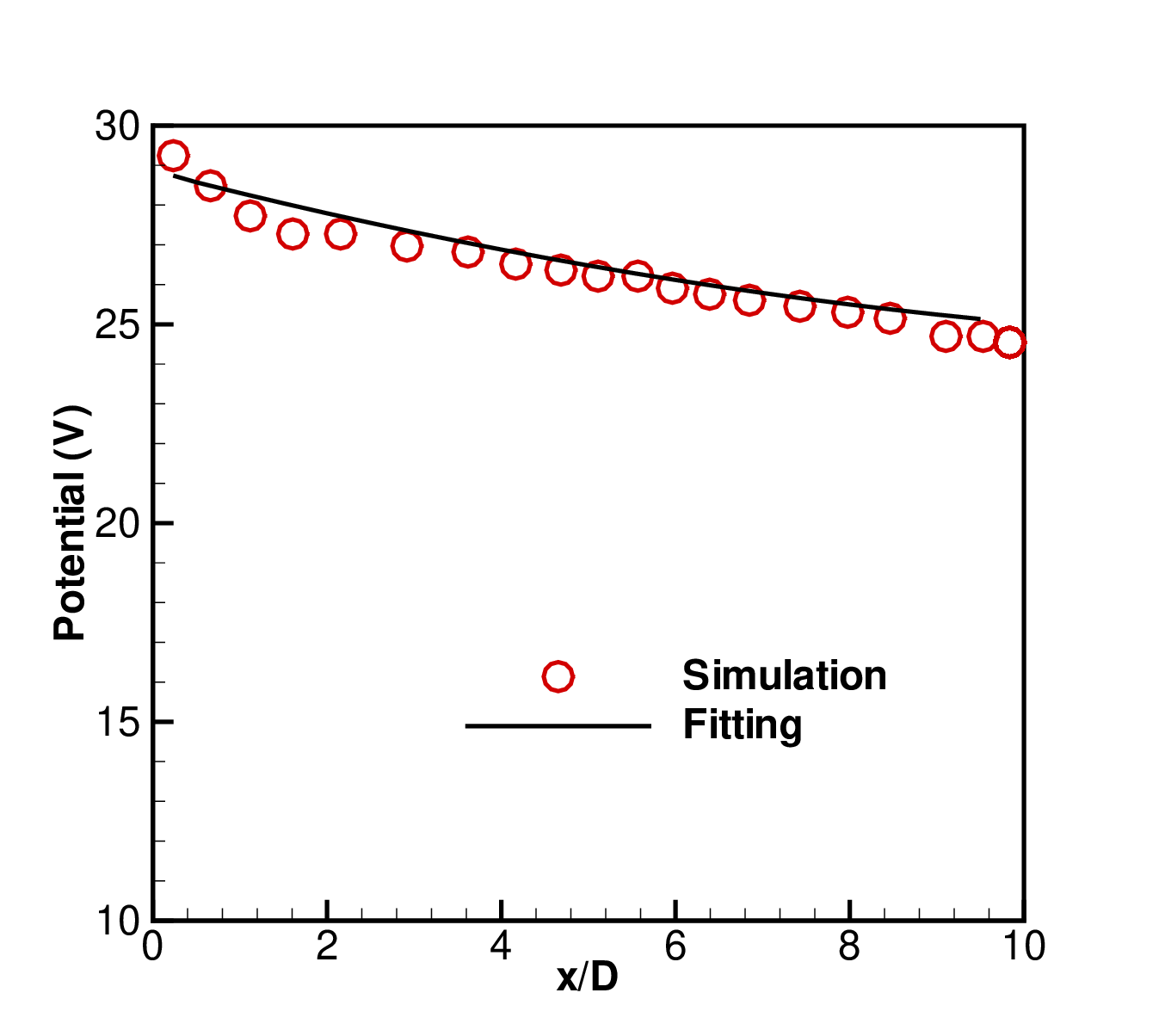}
      \caption{Curve-fitting results for the plasma potential profile along an H6 centerline. }
      \label{Fig:H6_Pot}
\end{figure}

The above two figures illustrate that from very limited simulation results we can construct the plasma potential and $T_e$ distributions which include two exponential functions. Is it possible to recover distributions over a wider range with two exponential functions, by using simulation results without any information on local properties? Figures \ref{Fig:H6_Te} and \ref{Fig:H6_Pot} present some hybrid numerical simulation results of plasma plume flows out an H6\cite{choi} with the hybrid simulation method. The cathode is placed at the H6 thruster center and aligned with the X-axis. As a result, the X-axis is a perfect streamline. These two figures include two smooth curve-fitting results for $T_e$s and plasma potentials, and each curve includes two exponential functions. In the simulations, at locations closer to the acceleration channel, the magnetic field leakage and the cathode effects may be important; hence the farfield curve is closer to Eqns. \ref{eqn:Tesoultion} and \ref{Eqn:potential}.

Some discussions are offered here before we conclude this sub-section. First, at farfield where all the assumptions are more reasonable, the $T_e$ can be computed from the plasma potential measurements, by using Eqn.\ref{eqn:Tesoultion}. This is due to the common factors in both formulas, and $\overrightarrow{a}$ and $b$ are constants through the whole flowfield. The profiles are smooth and the variations are mild at farfield. In fact, those smooth curves with mild variations are common features in plasma flows, for example, those $T_e$ profiles computed with the  Weakly Ionized Plasma (WIP) Model and the Magneto-Hydro-Dynamics (MHD) methods\cite{WIP}. Those curves can be conveniently approximated
with exponential functions obtained from this study.  Second, farfield properties can be obtained by extrapolations, based on the curve-fitting results. This is because Eqns. \ref{eqn:Tesoultion} and \ref{Eqn:potential} predict that $T_e$ or plasma potential profiles decrease as exponential functions. Experiments may offer limited amount of measurements, but Eqns.\ref{eqn:Tesoultion} and \ref{Eqn:potential} allow us to extrapolate. Third, Figs.\ref{Fig:singleBHT_cen_te_exp_detailed_fit} and \ref{Fig:H6_Te} illustrate that two curves are needed to represent the plasma $T_e$ profile in a plasma plume flow from a Hall effect thruster.  Plasma flows from the acceleration channel impinge at the thruster centerline in front of the thruster, creating the highest thermal temperature. $\lambda_1$ and $\lambda_2$ and coefficients are different at different sides of the peak because of the totally opposite trends there. Figs. \ref{Fig:singleBHT_cen_pot_exp_detailed_fit} and \ref{Fig:H6_Pot} illustrate probably one fitting-curve is sufficient to describe electron potential. However, we emphasize that these observations are based on simulation results for specific plasma plume flow out of a Hall thruster, and more investigations are needed in the future for other general dilute plasma flows. Fourth, the new $T_e$ and potential formulas contain more physics and factors than the simple Boltzmann relation; hence, these new formulas shall be relatively more reasonable and more widely applicable.

\subsection{Unsteady Flow Situation}
\label{subsec:unstady_validation}
There are more coefficients in Eqns.\ref{eqn:unsteady_Tesoultion} and \ref{eqn:unstady_pot}, and those parameters are functions of local properties. Hence, it is not practical to determine specific values by curve-fitting, as performed in the steady flow section.  Instead,  we can only demonstrate there exists possible trends in numerical simulation results. 

One hybrid three-dimensional simulation of  dilute plasma plume flows from  a Hall Effect Thruster BHT200 was performed, by using the same method and simulation packaged described for the steady flow situation. The simulation configurations, setups, and parameters are similar to the steady flow scenario.  The 1st stage of simulation was about a thruster and plume startup process, and the other was for a thruster shutting-down process. After a certain amount of time interval, the plasma potential and $T_e$ values at several points along the plume centerline are sampled and recorded. The following results are the property development histories which can demonstrate the trends. 

\begin{figure}[ht]
   \includegraphics[width=3.6in,height=2.9in]{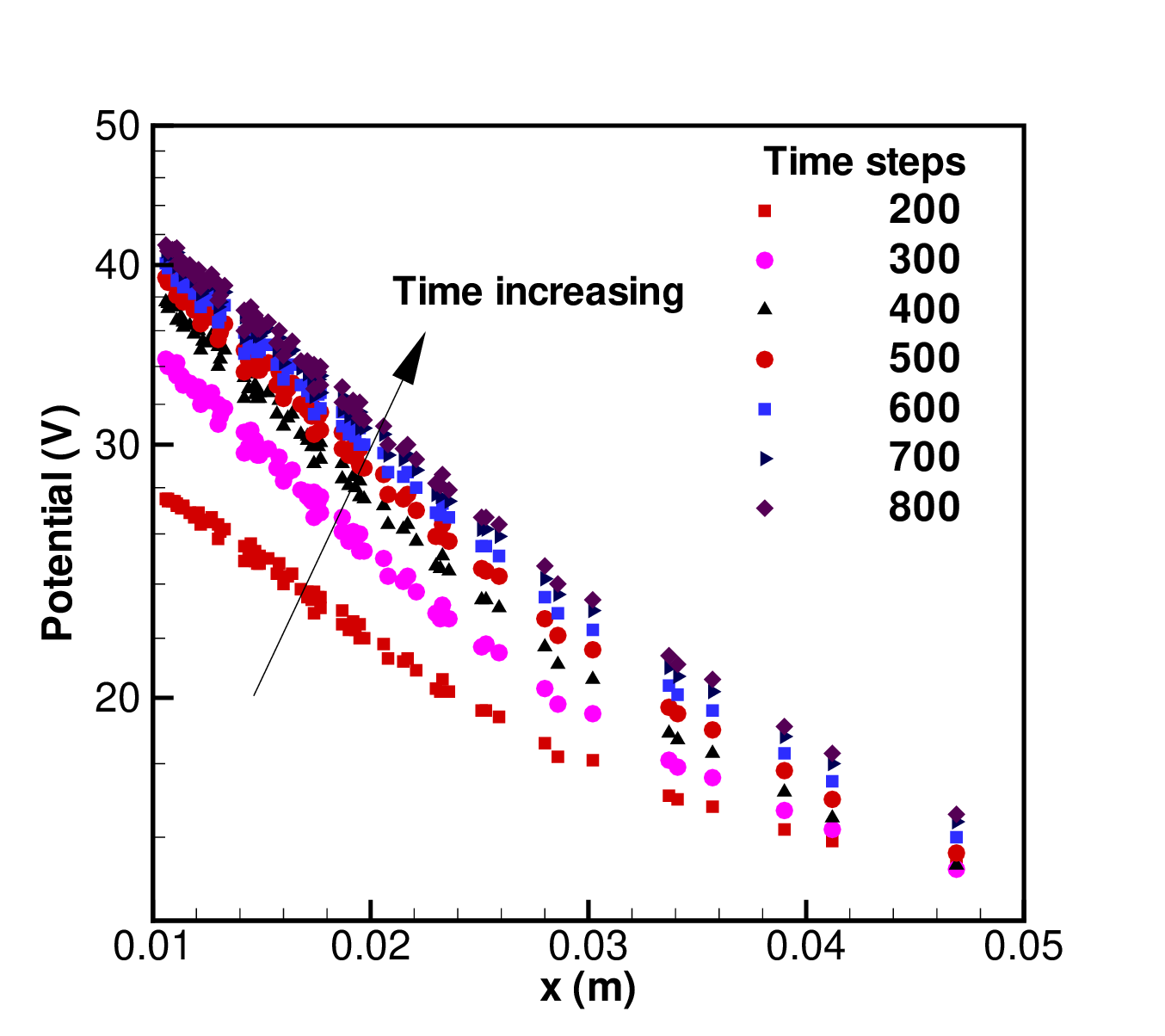}
      \caption{Centerline potential distribution in a startup process, the inner region.}
       \label{Fig:startup_phi_in}
\end{figure}

\begin{figure}[ht]
   \includegraphics[width=3.6in,height=2.9in]{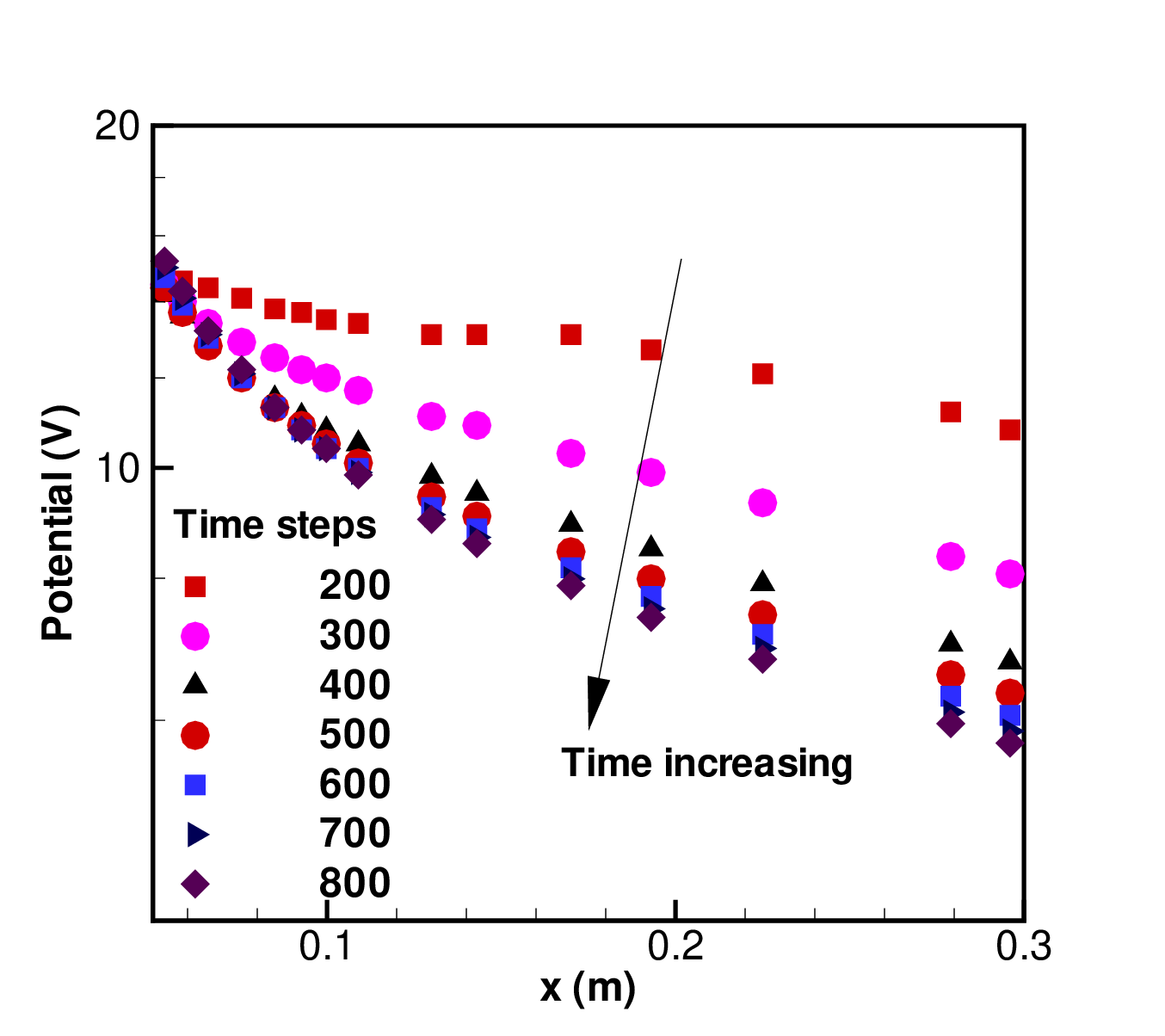}
      \caption{Centerline potential distribution in a startup process, the outer region. }
      \label{Fig:startup_phi_out}
\end{figure}

\begin{figure}[ht]
    \includegraphics[width=3.6in,height=2.9in]{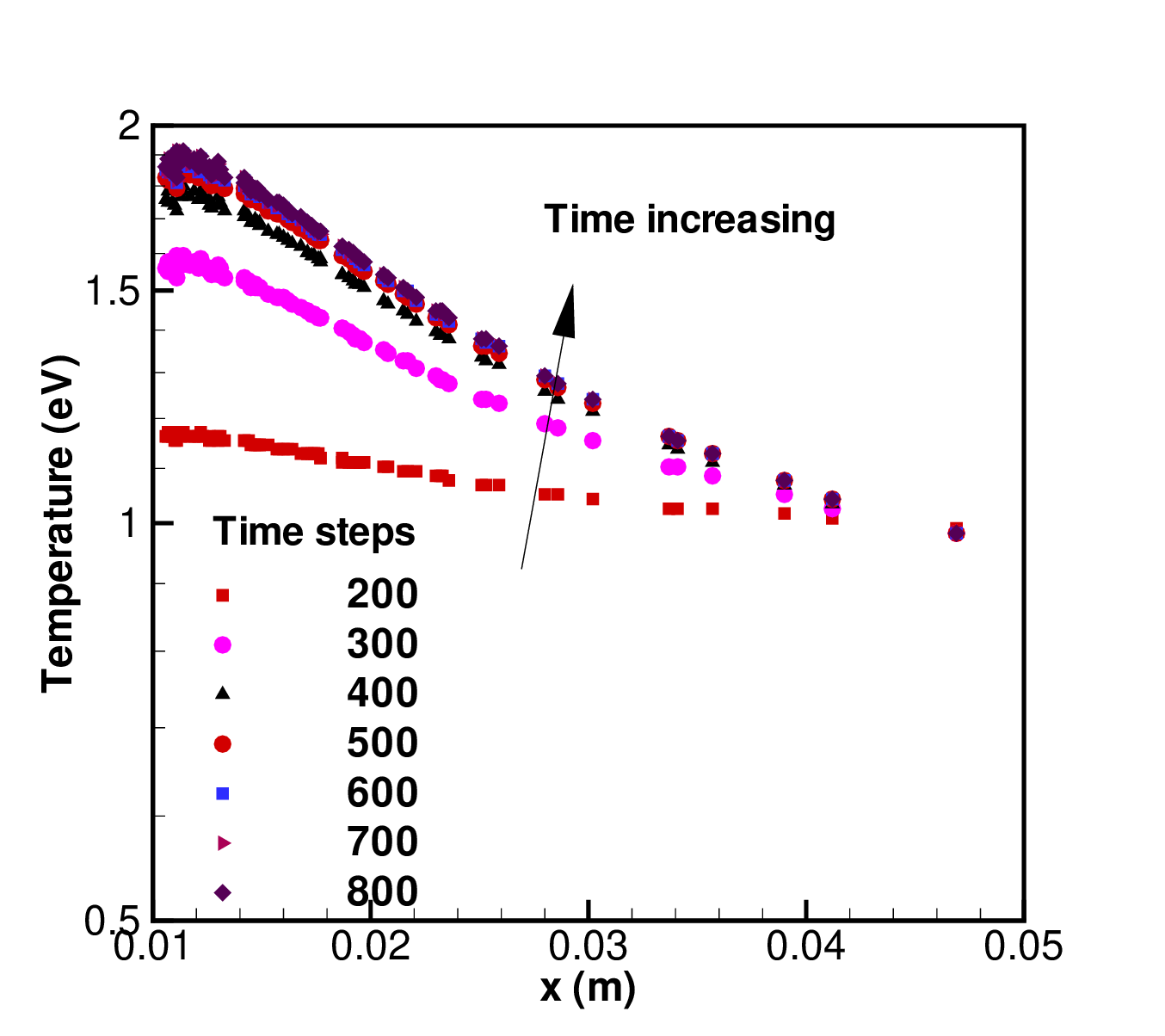}
      \caption{Centerline $T_e$ distribution in a startup process, the inner region.}
       \label{Fig:startup_te_in}
\end{figure}

\begin{figure}[ht]
 \includegraphics[width=3.6in,height=2.9in]{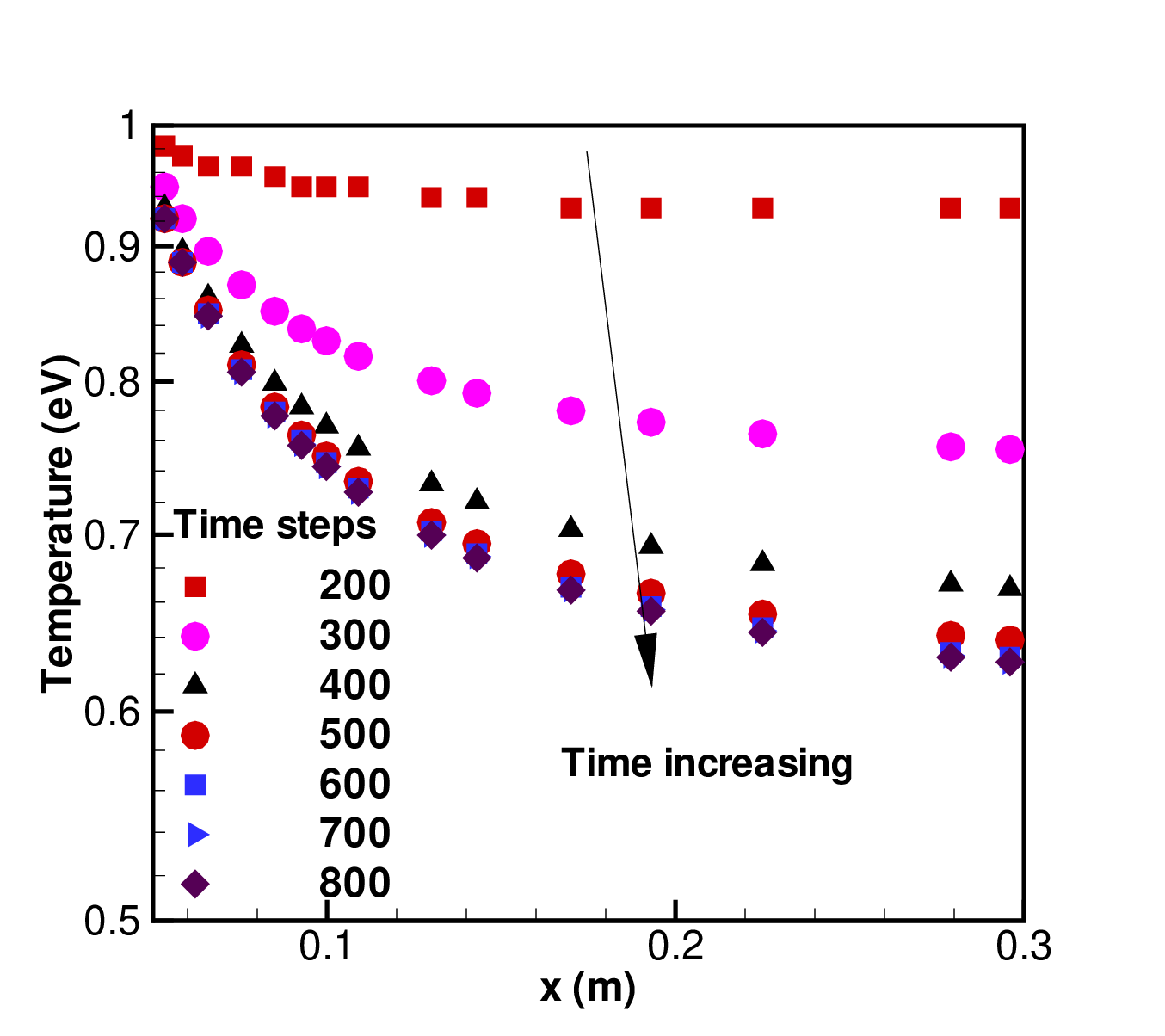}
      \caption{Centerline $T_e$ distribution in a startup process,  the outer region. }
      \label{Fig:startup_te_out}
\end{figure}

Figures \ref{Fig:startup_phi_in} - \ref{Fig:startup_te_out} are four figures presenting development histories for $T_e$s and potentials along the plume centerline for a plasma plume startup process, which may have an initial field inconsistent to the governing equations. This is the situation of cold plasma becomes  hotter.  The flowfield is initialized with specific conditions \cite{Cai0}. Once the simulation starts, the flowfield begins to develop and the electron properties respond to external changes. In these four figures, one specific symbols represent the values at a specific 
time. Figures \ref{Fig:startup_phi_in} and \ref{Fig:startup_phi_out} are plasma potential distributions in the inner and outer regions. These two regions are separated by the cathode-centerline impingement point, i.e., $x=0.05$ cm.  The inner region is close to the thruster exit plane, and the latter is beyond the impingement point, and in the downstream region. The plotted curves are sampled every certain steps. The Y-axis is  plotted in log scales; hence, quasi-linear relations between the X- and Y- axis  appear, as predicted by Eqn.\ref{eqn:unstady_pot},  and at the end of development, an asymptote profile appear.  These two groups of curves have opposite trends and are related to the initial fields.  Figures \ref{Fig:startup_te_in} and \ref{Fig:startup_te_out} are the corresponding centerline $T_e$ profiles, and they share similar patterns with the potential fields, as Eqn.\ref{eqn:unsteady_Tesoultion} predicted. Both the potential and $T_e$ equations include exponential functions. The corresponding asymptote profile are the final steady state values.   For those four figures, the profiles become smoother at the end of the startup stage, and the end of shutting down processes, because the local plasma properties are fully developed with constant plasma constant.    

\begin{figure}[ht]
  \includegraphics[width=3.6in,height=2.9in]{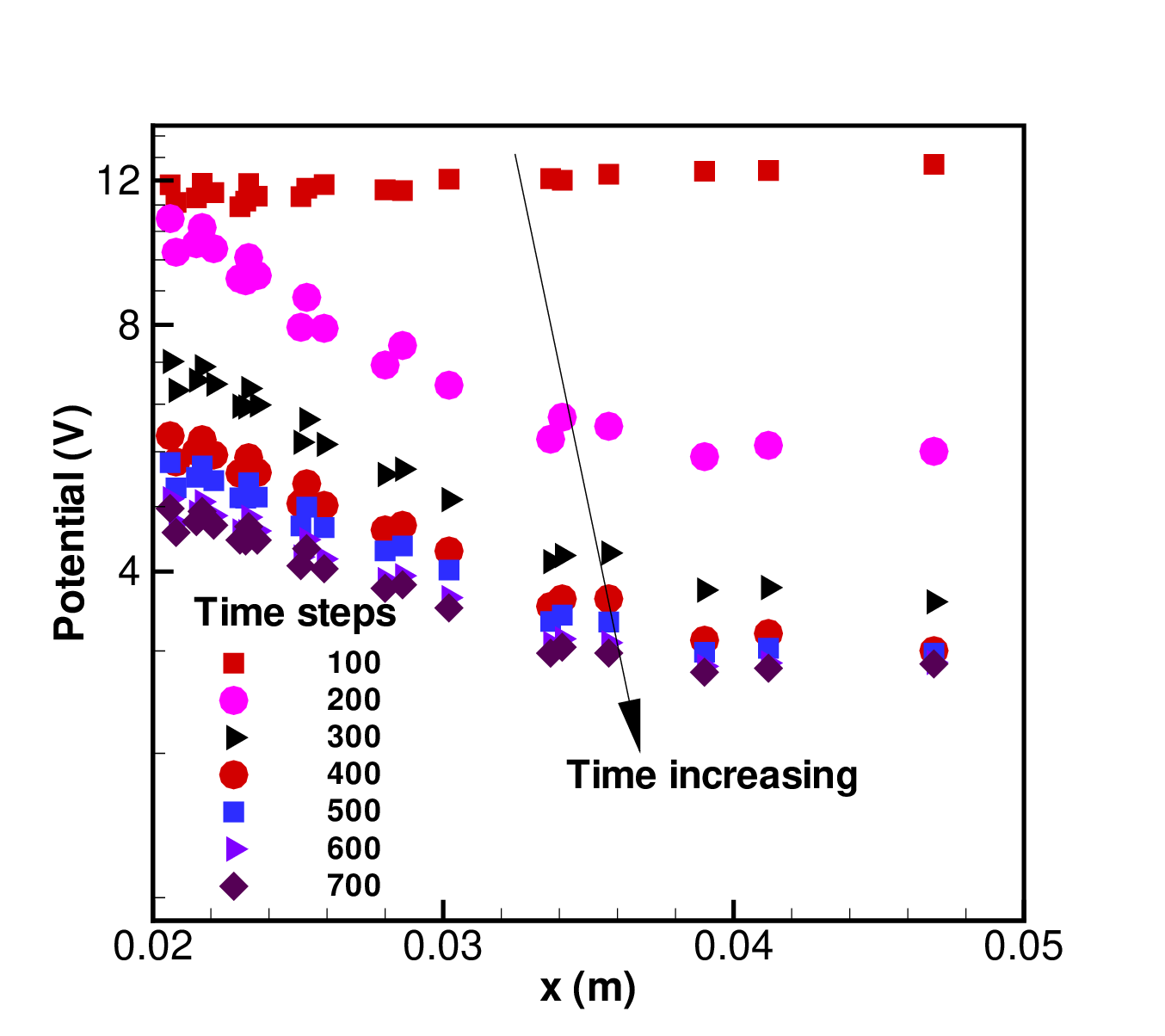}
      \caption{Centerline potential distribution in a shutdown process,  the inner region.}
       \label{Fig:off_phi_in}
\end{figure}

\begin{figure}[ht]
\includegraphics[width=3.6in,height=2.9in]{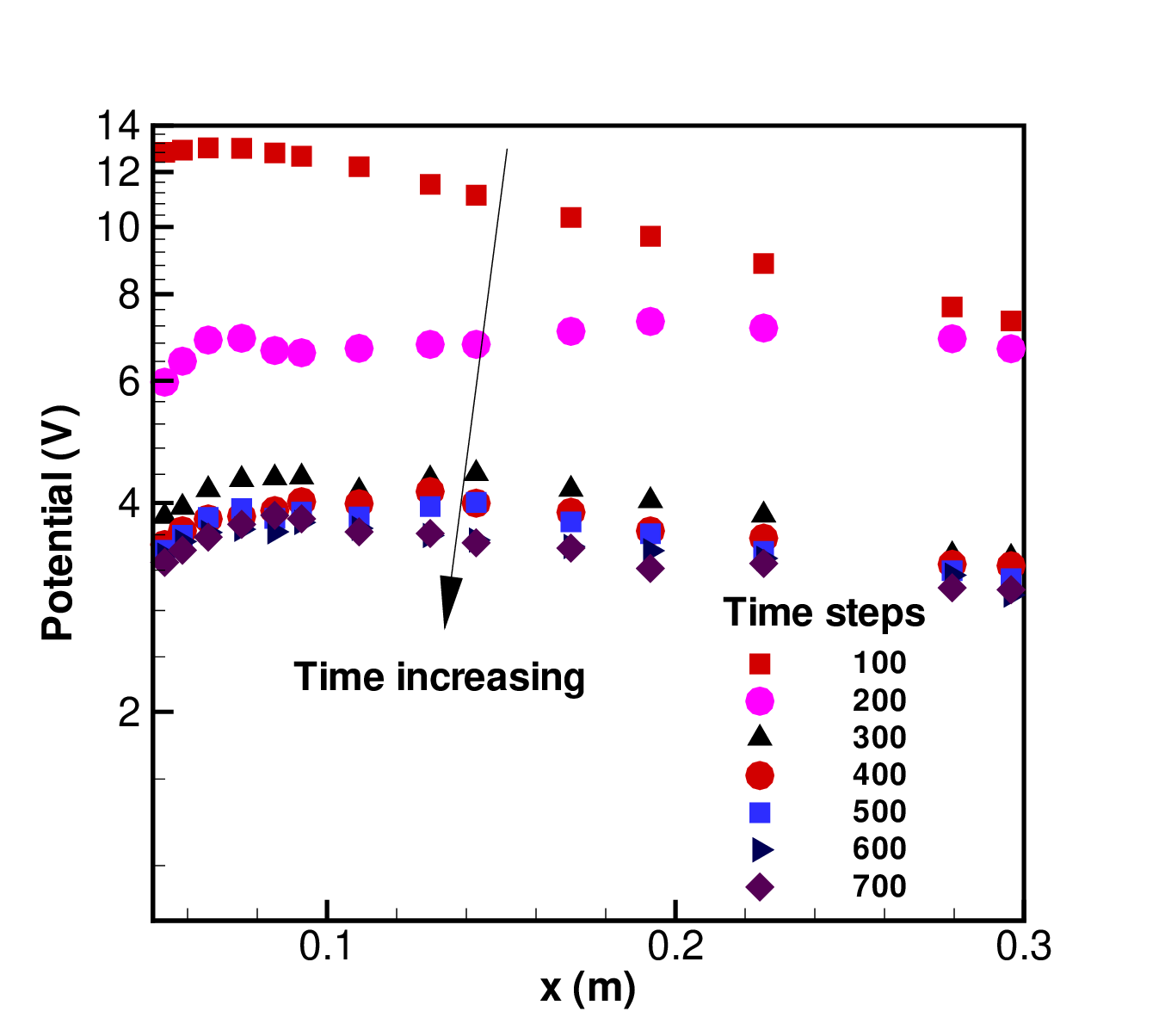}
      \caption{Centerline potential distribution in a shutdown process, the outer region. }
      \label{Fig:off_phi_out}
\end{figure}

\begin{figure}[ht]
  \includegraphics[width=3.6in,height=2.9in]{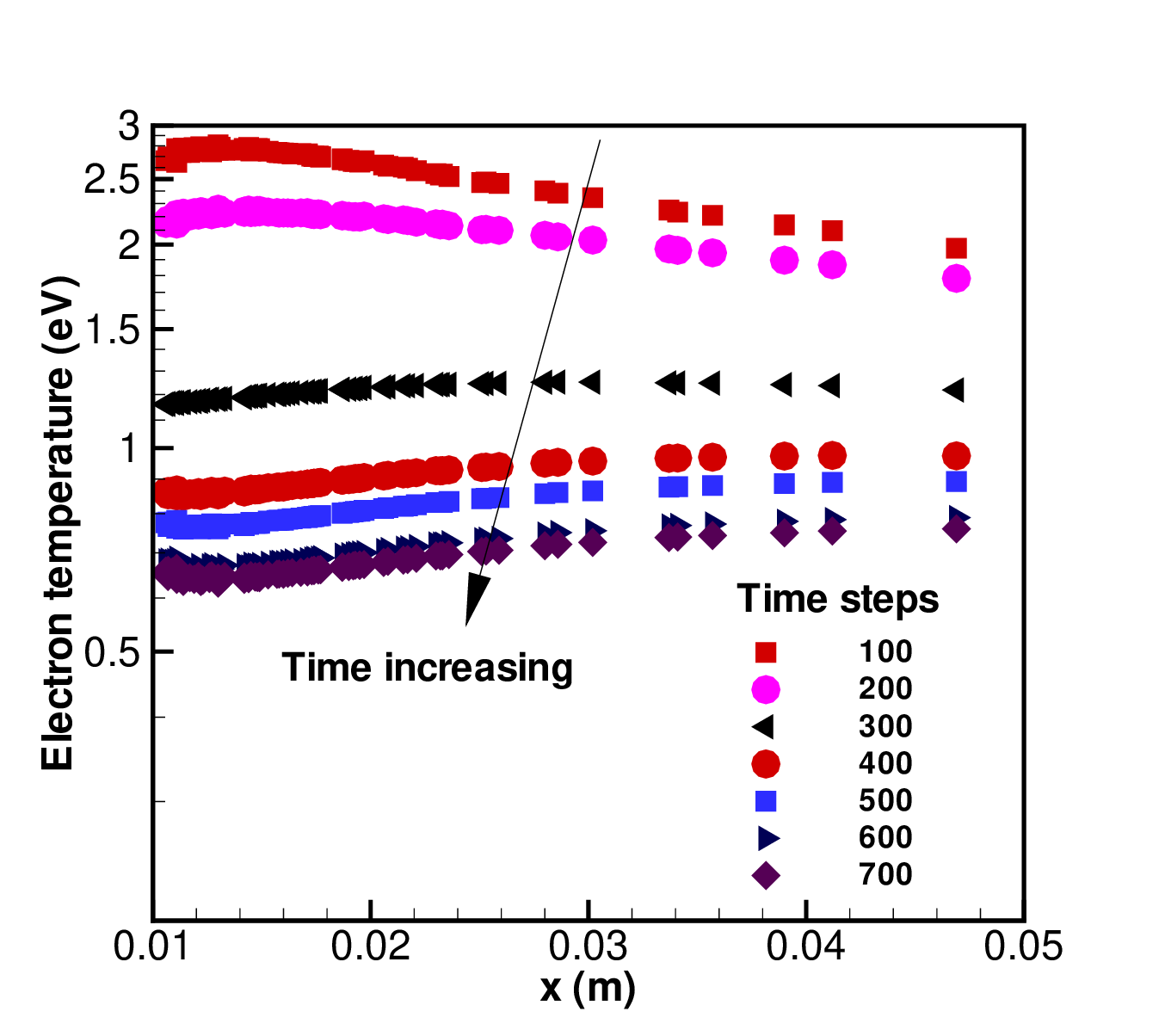}
      \caption{Centerline $T_e$ distribution in a shutdown process,  the inner region.}
       \label{Fig:shutdown_te_in2}
\end{figure}

\begin{figure}[ht]
 \includegraphics[width=3.6in,height=2.9in]{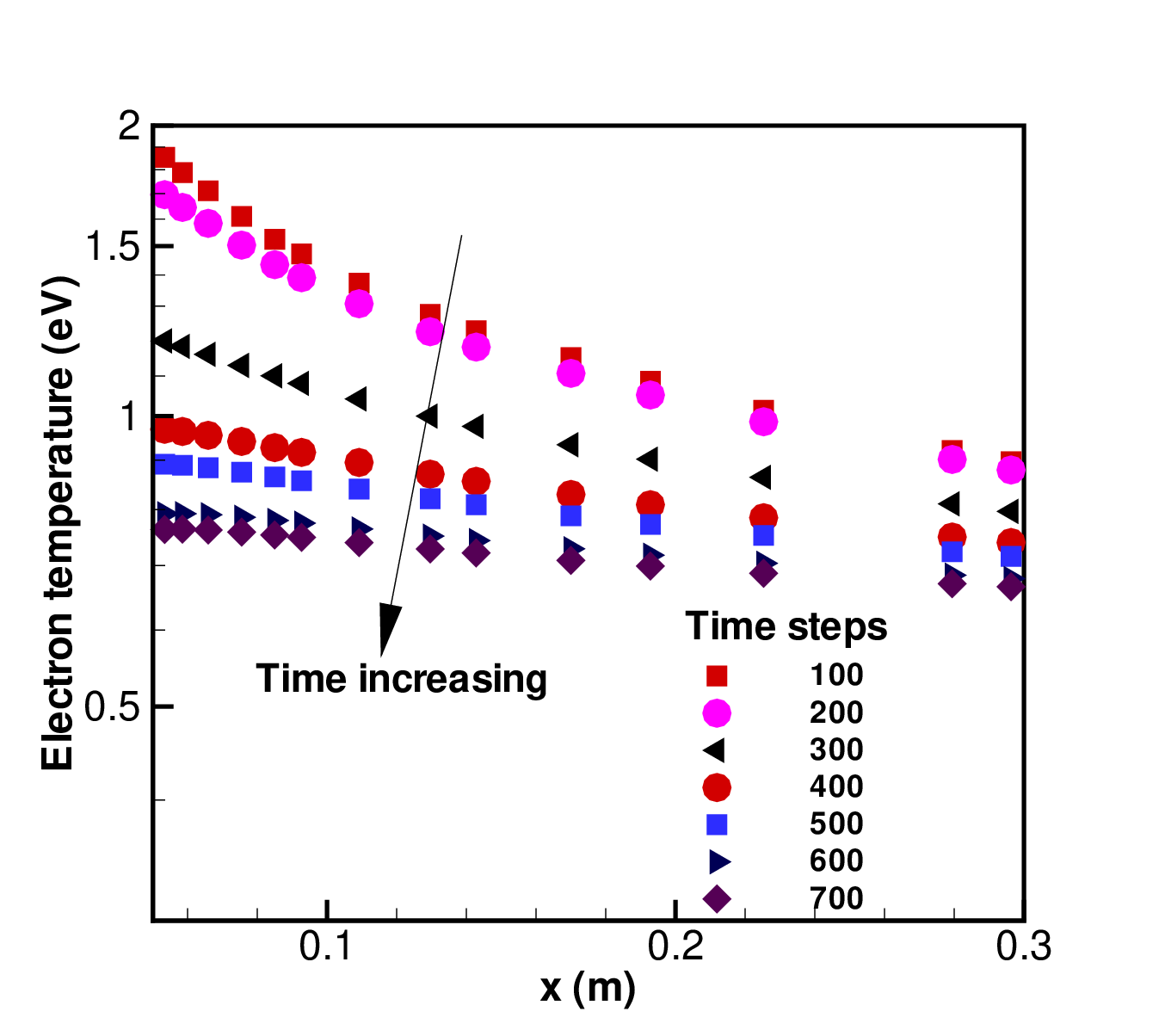}
      \caption{Centerline $T_e$ distribution in a shutdown process, the outer region.}
      \label{Fig:shutdown_te_out2}
\end{figure}

Figures \ref{Fig:off_phi_in} - \ref{Fig:shutdown_te_out2}  present development histories for plasma potential and $T_e$ in a shutdown process, i.e., the hot plasma flows become cooler. Figures \ref{Fig:off_phi_in} and \ref{Fig:off_phi_out} are for a plasma potential development. As shown, in the inner and outer regions, potentials monotonically decrease, and each curve follows a pattern of  quasi-linear relation along the streamline direction, indicating exponential relations exist in the true distributions.  These two figures also show that there are asymptotes for the unsteady development. Figure \ref{Fig:shutdown_te_in2} and \ref{Fig:shutdown_te_out2} show the corresponding $T_e$ distributions during the same shutting down process,   Similar patterns as the plasma potentials are observed.  At the end of development, i.e., closer to  800 time steps, the potential and  $T_e$s converges, and the distances among them approach to constant values.  At the final steady state, there are still none-zero distributions along the centerline, because a solution to three Poisson's equations \cite{Cai0} need to be satisfied with non-uniform boundary conditions.

\begin{figure}[ht]
    \includegraphics[width=3.6in,height=2.9in]{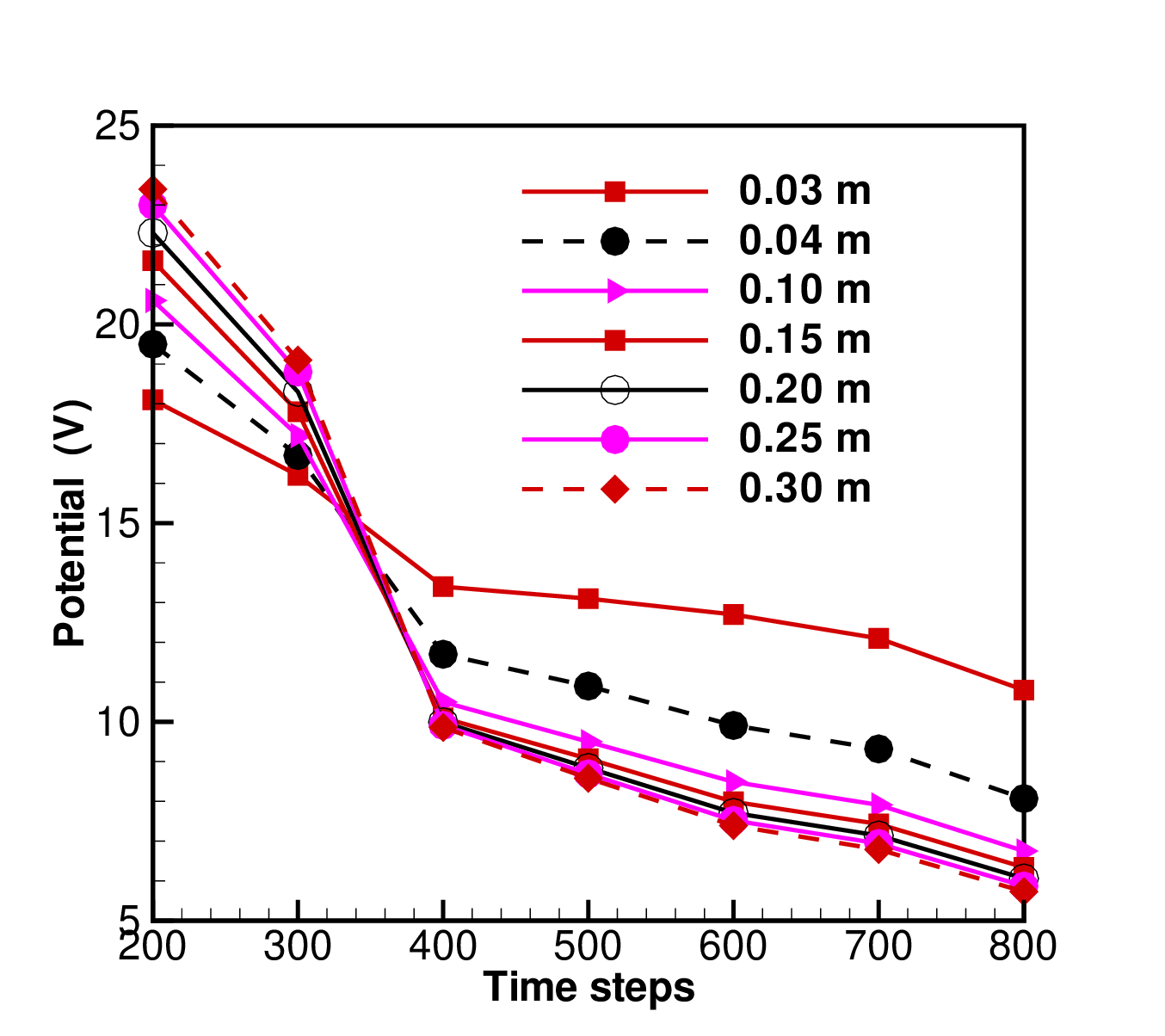}
      \caption{Plasma potential development history in a startup process.}
       \label{Fig:startup_history_phi}
\end{figure}

\begin{figure}[ht]
    \includegraphics[width=3.6in,height=2.9in]{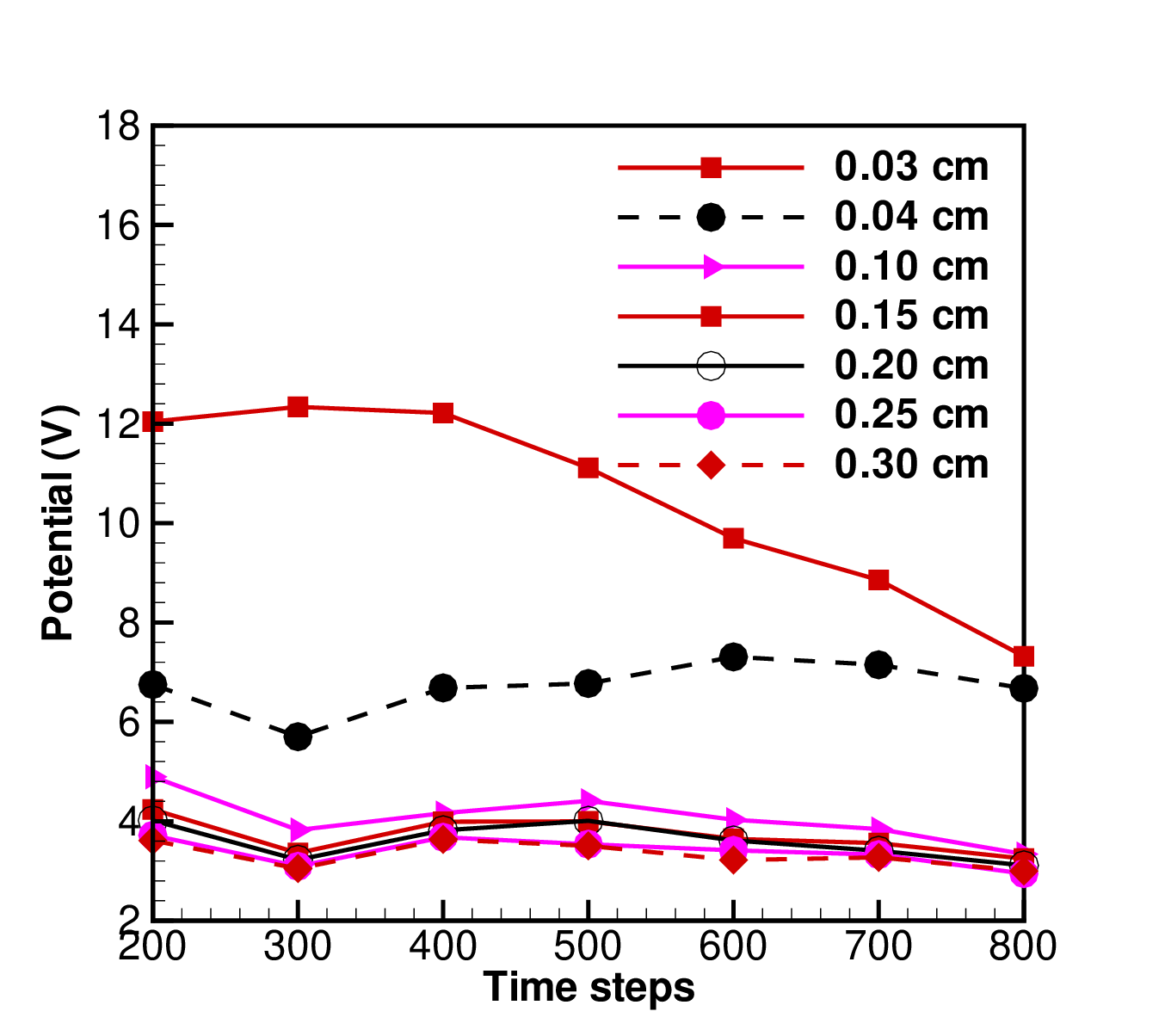}
      \caption{Plasma potential development history in a shutdown process.}
       \label{Fig:off_history_phi}
\end{figure}

\begin{figure}[ht]
 \includegraphics[width=3.6in,height=2.9in]{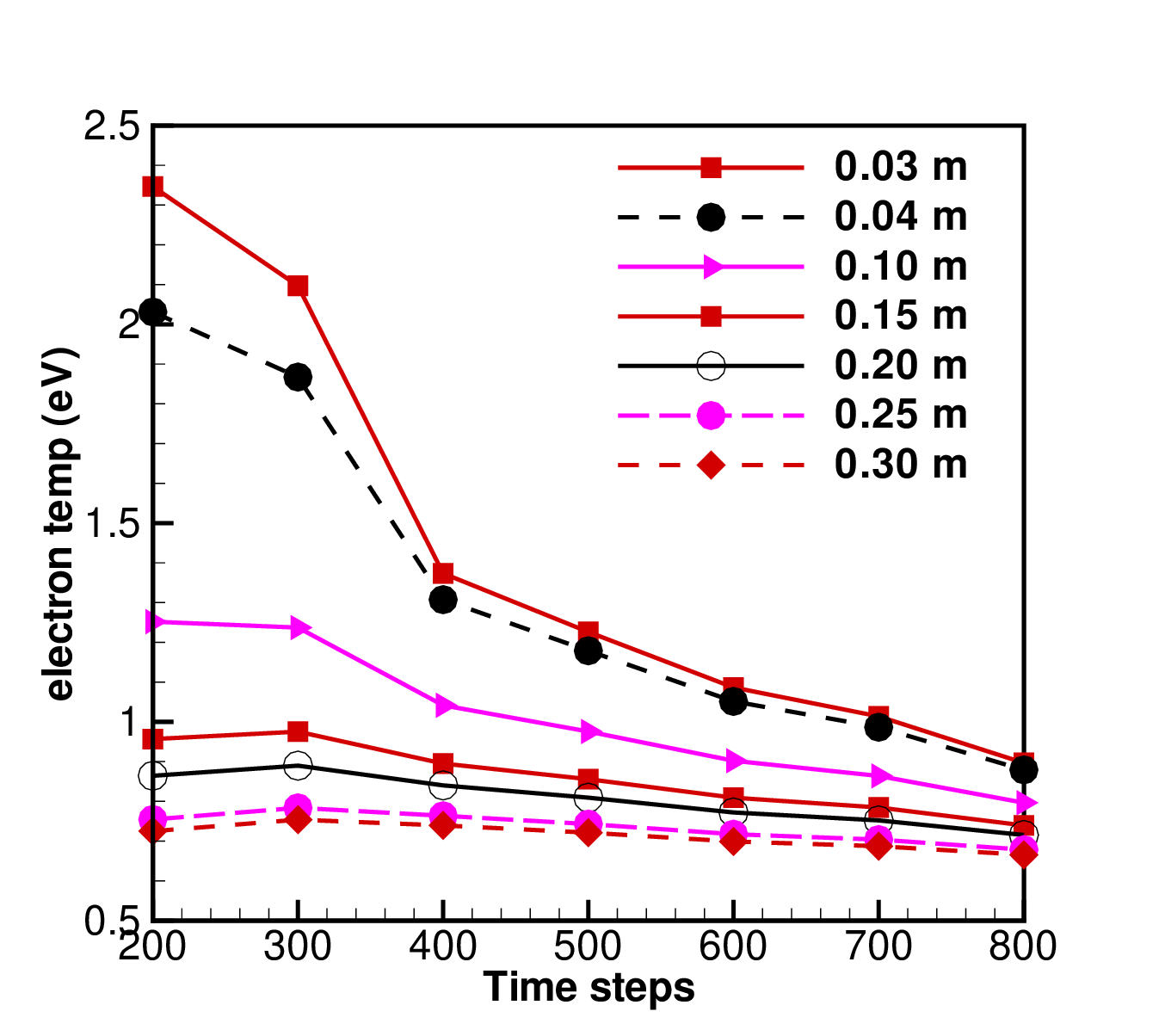}
      \caption{$T_e$ development history in a startup process.}
      \label{Fig:startup_history_Te}
\end{figure}

\begin{figure}[ht]
    \includegraphics[width=3.6in,height=2.9in]{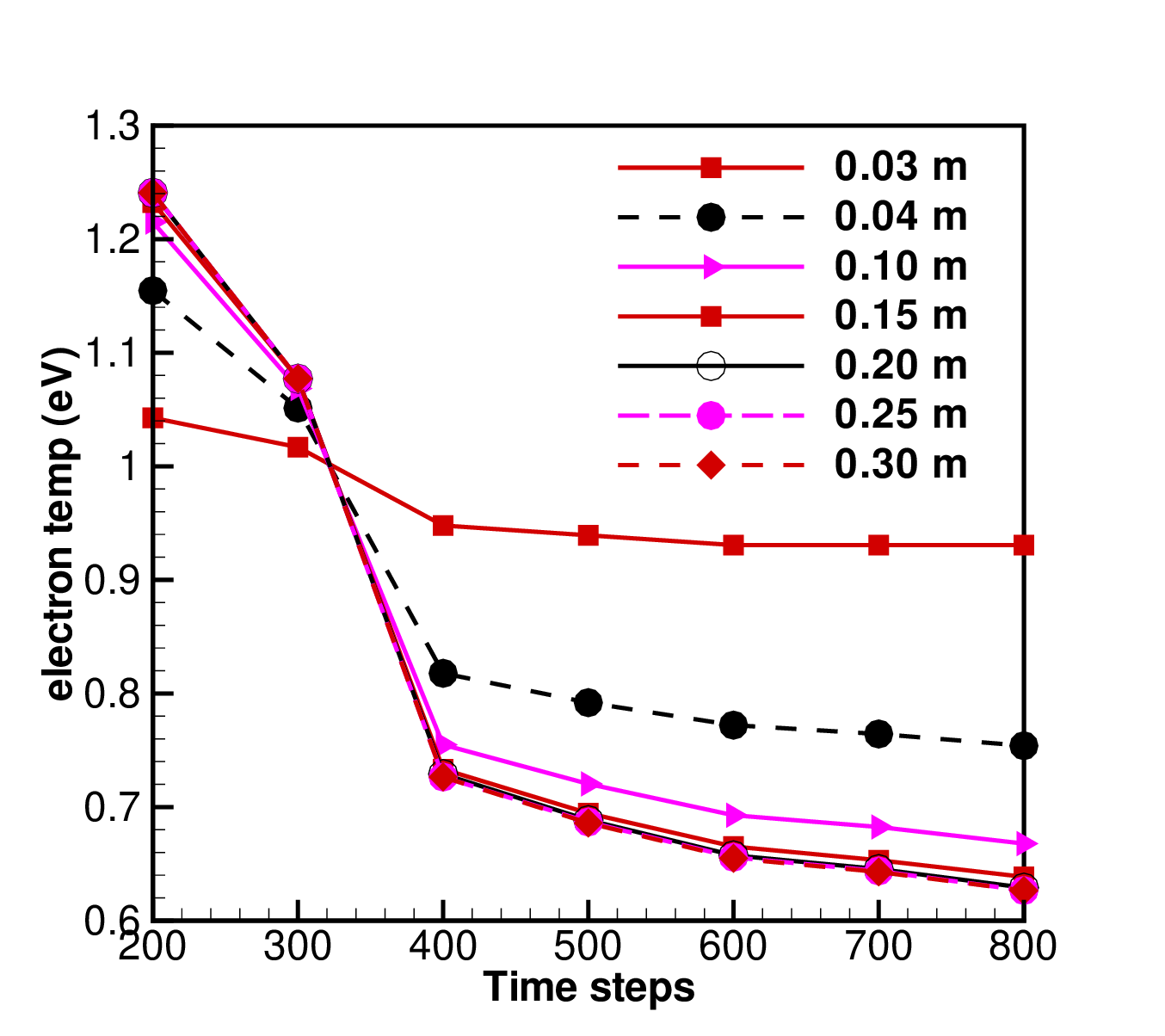}
      \caption{$T_e$ development history in a shutdown process.}
      \label{Fig:off_history_Te}
\end{figure}

Figures \ref{Fig:startup_history_phi} - \ref{Fig:off_history_Te}   are sampled potential and $T_e$ profiles at specific centerline locations for a startup and a shutdown process, correspondingly.  Figures \ref{Fig:startup_history_phi} and  \ref{Fig:off_history_phi} are potential results  which illustrate the unsteady exponential development history. In these two figures, the Y-axis is plotted in a linear scale. In the startup process, those profiles which are similar to exponential functions are evident, with sharp decreasing patterns. In the shut down process, the profiles are relatively flat, essentially reaching the asymptotes or the ambient conditions. Figures \ref{Fig:startup_history_Te} and \ref{Fig:off_history_Te} show the corresponding $T_e$ development history at the same locations, similar exponential function patterns are evident.  In these four figures,  the curves  for the inner region spots, $x=0.03, 0.04 <x_p=0.05$ cm, have rapid changes but all the other curves have mid development histories. This is because these in the inner region, the gradients are relatively larger and those assumptions for the correlation derivations are poorly satisfied than the outer region. 

Some discussions are offered here to end this sub-section. First, in the outer region,  the assumptions are more reasonable; the space gradients in plasma properties are mild, and the profiles there are smoother, and are closer to each other. It is possible to extrapolate farfield properties, based on the curve-fitting results. This is because they decrease exponentially. Experiments may offer limited measurements, but results from this work can offer more information. Secondly, those figures illustrate that two curves are needed to represent the plasma $T_e$ profile in a plasma plume flow from a Hall effect thruster.  The plasma plume flows from the annular acceleration channel and the cathode impinge at the thruster centerline in front of the thruster, creating the highest electron thermal temperature. Related coefficients in those formulas vary at different sides of the impingement point,  and  may have different development trends. However, it shall be emphasized that these observations are based on simulation results for specific plasma plume flow out of a Hall effect thruster, and more investigations are needed for other unsteady, dilute, cold plasma flows.

\section{Conclusions}
\label{sec:final_conclusions}

Numerical simulations of plasma plume flows from EP devices were performed with the hybrid method, where electrons are treated as fluids, but heavy ions and neutrals are simulated with the DSMC and PIC method. The simulation results demonstrate clear exponential patterns along the electron streamlines for both steady and unsteady plasma flows.  Even though more test work is needed, the results are generally encouraging.




\begin{thebibliography}{17}
\expandafter\ifx\csname
natexlab\endcsname\relax\def\natexlab#1{#1}\fi
\expandafter\ifx\csname bibnamefont\endcsname\relax
  \def\bibnamefont#1{#1}\fi
\expandafter\ifx\csname bibfnamefont\endcsname\relax
  \def\bibfnamefont#1{#1}\fi
\expandafter\ifx\csname citenamefont\endcsname\relax
  \def\citenamefont#1{#1}\fi
\expandafter\ifx\csname url\endcsname\relax
  \def\url#1{\texttt{#1}}\fi
\expandafter\ifx\csname urlprefix\endcsname\relax\def\urlprefix{URL
}\fi \providecommand{\bibinfo}[2]{#2}
\providecommand{\eprint}[2][]{\url{#2}}  

\bibitem{PIC}
Birdsall, C.K., and Langdon,  A.B., {\it Plasma Physics Via Computer Simulation}, Institute of Physics, 1991.

\bibitem{chen}
Chen, F.F., {\em Introduction to Plasma Physics And Controlled  Fusion}, 2nd edition, Springer, Haryana, India, 2006.


\bibitem{yim}
Boyd, I. D., and Yim, J.T., ``Modelling Of The Near Field Plume Of A Hall Thruster,'' {\em Journal of Applied Physics}, Vol.95, 2004, pp.4575-4584.

\bibitem{DSMC}
Bird, G.A.,  {\it Molecular Gas Dynamics and The Direct Simulation of Gas Flows}, Clarendon Press, 1994.

\bibitem{Review}
Ahedo, E., ``Using Electron Fluid Models to Analyze Plasma Thruster Discharges,'' {\em Journal of Electric Propulsion}, (2023) 2:2. https://doi.org/10.1007/s44205-022-00035-6. 

\bibitem{two}
Sentis, R., Paillard, D., Baranger, C., and Seytor, P., ``Modelling And Numerical Simulation Of Plasma Flows With Two-Fluid Mixing,'' {\em European Journal of Mechanics B/Fluids}, Vol.30, 2011, pp.252-1258.

\bibitem{aristov}
Aristov, V.V., {\em Direct Methods for Solving the Boltzmann Equation And Study of Non-equilibrium Flows}, Kluwer Academic Publishers, 2001, ISBN 1-4020-0388-9.

\bibitem{Cooke}
Cooke, D., and Katz, I., ``TSS-1R Electron Currents: Magnetic Limited Collection From A Heated Presheath,'' {\em Geophysical Research Letters}, Vol.25, No.5, pp.736--756, March 1,  1998.

\bibitem{pekker}
Pekker, L, and Hussary, N., ``Effect Of Boundary Conditions On The Heat Flux To The Wall In Two-Temperature Modelling of Thermal Plasmas,'' {\em Journal of Physics D: Applied Physics}, Vol.45, 2012, 465206.

\bibitem{weifz}
Wei, F., Wang, H.X., Murphy, A.B., Sun, W.P., and Liu, Y., ``Numerical Modelling Of The Nonequilibrium Expansion Process Of Argon Plasma Flow Through A Nozzle,'' {\em Journal of Physics D: Applied Physics}, Vol.46, 2013, pp.505205.







\bibitem{Cai0}
Cai, C., ``Numerical Investigations on Plasma Plume Flows From a Cluster Of Electric Propulsion Devices'', {\it Aerospace Science \& Technology}, Vol.41, pp.134-143.  doi:10.1016/j.ast.2014.12.018.

\bibitem{Cai1}
Cai, S., and Cai, C., ``Two Relations in Steady and Dilute Plasma Flows,''  31$^{st}$ International Rarefied Gas-dynamic Symposium, Glasgow, Scotland, July, 2018.  AIP Conference Proceedings, Vol.2132, No.1, pp.040001, 2019. https://doi.org/10.1063/1.5119532.

\bibitem{Cai2}
Cai, C., and Cooke, D.,  ``A Simple Model for Electron Temperature In Dilute Plasma Flows,'' {\it Physics of Plasmas}, Vol.23, 103513, 2016. doi:http://dx.doi.org/10.1063/1.4965229.

\bibitem{unsteady}
K. Zhang, K., C. Cai, C., and Cooke, D., ``Simple Relations for Electron Temperature and Potential in Dilute Cold Plasma Flows,'' {\em J. Plas. Phys.}, Vol.25, 012128, 2018. https://doi.org/10.1063/1.5010768.

\bibitem{Mitcher}
Mitcher, M., and Kruger, C.H.,  {\it Partially Ionized Gases}, Wiley, 1973.

\bibitem{choi}
M. Choi, {\em Improved  Hall Thruster Plume Simulation by Including  Magnetic Field Effects},  Ph.D. dissertation, Dept. of Aerospace Engineering, the University of Michigan,  Ann Arbor, 2017. 

\bibitem{WIP}
Siddanathi, S.L., Westerberg, L.G., Akerstedt, H.O., Wiinikka, H., and Sepman, A., ``Computational Modeling of a Plasma Torch Using Single-Fluid and Two-Fluid Modelling Approaches,''  2022 IEEE International Conference on Plasma Science (ICOPS), doi: 10.1109/ICOPS45751.2022, 22-26 May 2022. 

\end{thebibliography}
\end{document}